# The discharge characteristics of the DUHOCAMIS with a high magnetic bottle-shaped field*

FU Dong-Po (付东坡)[1]   ZHAO Wei-Jiang (赵渭江)[1,1)]   GUO Peng (郭鹏)[1]   ZHU Kun (朱昆)[1]   WANG Jing-Hui (王景辉)[2]   HUA Jing-Shan (华景山)[1]   REN Xiao-Tang (任晓堂)[1]   XUE Jian-Ming (薛建明)[1]   ZHAO Hong-Wei (赵红卫)[3]   Liu Ke-Xin (刘克新)[1]

[1] Institute of Heavy Ion Physics & State Key Laboratory of Nuclear Physics and Technology, Peking University, Beijing 100871, China
[2] The Ohio State University, Nuclear Engineering Program, Columbus, OH 43210, USA
[3] Institute of Modern Physics, Chinese Academy of Science, Lanzhou 730000, China

**Abstract**

For the purpose to produce high intensity, multiply charged metal ion beams, the DUHOCAMIS (dual hollow cathode ion source for metal ions) was derived from the hot cathode Penning ion source combined with the hollow cathode sputtering experiments in 2007. It was interesting to investigate the behavior of this discharge geometry in a stronger magnetic bottle-shaped field. So a new test bench for DUHOCAMIS with a high magnetic bottle-shaped field up to 0.6 T has been set up at Peking University, on which have been made primary experiments in connection with discharge characteristics of the source. The experiments with magnetic fields from 0.13 T to 0.52 T have shown that the magnetic flux densities are very sensitive to the discharge behavior: discharge curves and ion spectra. It has been found that the slope of discharge curves in a very wide range can be controlled by changing the magnetic field as well as regulated by adjusting cathode heating power. On the other hand, by comparison of discharge curves between dual hollow cathode discharge (DHCD) mode and PIG discharge mode, it was found a much stronger magnetic effect occurred on DHCD mode. In this paper, the new test bench with ion source structure is described in detail; and main experimental results are presented and discussed, including the effects of cathode heating power and magnetic flux density on discharge characteristics, also the ion spectra. The effects of the magnetic field on the source operating are emphasized, and a unique behavior of the DUHOCAMIS operating in the high magnetic field is expected and discussed especially.

## 1. Introduction

Metal ion beams have been widely used in industrial process such as coating, etching and implantation for material surface modifications, and there has been an ever increasing demand for high intensity, multi-charged metal ions for pure science and accelerator applications [1-4]. Many kinds of ion source can produce metal ions, such as, Penning Ion Gauge (PIG), Electron cyclotron resonance (ECR), Metal vapor vacuum arc (MEVVA), Electron beam ion source (EBIS) and Laser ion source (LIS), etc. All of them have their own advantages and limitations in producing metal ion beams in case of ion species, current density, charge state, beam size, operation mode, and duty cycles and so on. Considering the high utilization ratio of metal materials and the high universals of a metal ion source, the DUHOCAMIS (Dual Hollow Cathode Ion Source for Metal Ion Beams) was presented in 2007 [5,6], in which a tubular sputter co-cathode united in a magnetic bottle-shaped field has been developed based on the hot cathode PIG. This kind of ion source would be beneficial to obtain high metal ion current, high charge state ions, high plasma stability, and high material utilization ratio of various metal species. So this source would be expected as a high universal and very convenient metal ion source to produce metal ion beams in a wide range of ion species, ion energy, ion current; it also can deliver gas ion beams when working at the modified-PIG mode, which is distinguished from the traditional PIG [1, 6] in two points: (1) using the bottle-shaped magnetic field, and (2) the hollow anode being consisting of three coaxial parts with two gaps. But, there was no more knowledge and experience about this kind of source before. It was much interesting to investigate the behavior of this discharge geometry in a stronger magnetic bottle-shaped field. So a new test bench for DUHOCAMIS with a high magnetic bottle-shaped field up to 0.6 T has been set up at Institute of Heavy Ion Physics, Peking University. Up to now, we have done a series of experiments related to the discharge properties of the source, which was operated in a wide range of bottle-shaped magnetic field with different cathode heating powers, Argon flows, and pulse modes for arc power supply. Some ion spectra were measured by an analyzing magnet. To understand the special features of dual hollow cathode discharge (DHCD) mode, we compared discharge behavior between DHCD mode and PIG discharge mode under the same ion source-geometry and comparable discharge conditions.

In this paper, we will describe the ion source structure and the new test bench in detail, also present and discuss a series of experimental results in connection with the discharge characteristics of the source, including the effects of the cathode heating power and the magnetic flux density, as well as the ion spectra. On the other hand, it will be much interested in comparison of discharge characteristics between DHCD and PIG discharge mode. The effects of the magnetic field on the source operating are emphasized and discussed especially. The unique properties of the DUHOCAMIS would be expected and discussed for DHCD mode operating in a higher magnetic bottle-shaped field higher than about 0.4 T.

* Supported by National Natural Science Foundation of China (11105008, 10775011)
1) E-mail: wjzhao@pku.edu.cn

## 2. Experimental setup
### 2.1 Ion source structure

The discharge chamber of DUHOCAMIS is schematically shown in Fig. 1 together with its outer circuit. It consists of a hollow tubular sputter cathode (4) with one anode (3, 5) and one cathode (2, 6) on each side, and the whole structure is coaxially aligned on the axis of a bottle-shaped magnetic field. The H-type magnet (7) was used to generate a magnetic bottle-shaped field with a high flux density in a magnet gap of 20 cm, which was designed by the aid of Computer Simulation Technology (CST) software package. The maxima of the magnetic flux density $B_y$ can reach to about 0.6 T at the center of the magnet; also, the magnetic mirror ratio, i.e., the ratio of the field near the pole surface to the median plane, can be as high as 2.0. The high magnetic field combines with the high mirror ratio may confine better the electrons in the discharge chamber to increase the ionization probability, the plasma density and the plasma stability, thus to produce high ion current and highly charged ions.

During operations, electrons emitted from the filament (1) bombard the indirectly heated cathode (2) to form thermal-emission electrons, and these thermal electrons will be accelerated and scattered into the discharge chamber. The use of indirectly heated cathode is essential for multiply charged ion production, since it makes the accurate control and maintaining of the discharge parameters possible [7].

A noticeable feature of this configuration is that the source could be changed operating mode between DHCD and PIG discharge mode easily, when the switch K is turned on the contact D (i.e., the potential of the sputter cathode is the same as the cathode) [8, 9] or P (i.e., the potential of the sputter cathode is the same as the anode) [10, 11], respectively.

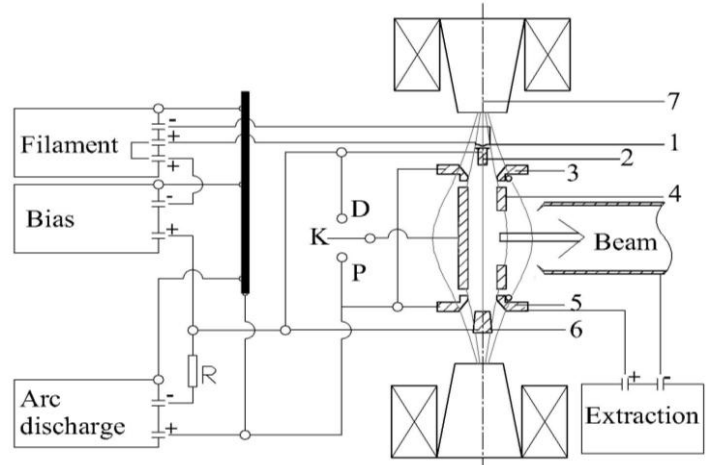

Fig.1. Scheme of the ion source structure: (1) filament, (2) heated cathode, (3 and 5) anodes, (4) tubular sputter cathode, (6) cold cathode and (7) magnet.

### 2.2 Extraction system and test bench

The test bench for the ion source is shown in Fig. 2, which mainly includes an extraction system, an analyzing magnet, and a target chamber. The extraction system composed of a triode extraction system (2-4), a magnetic shield (6) and an electrostatic deflector (5) [12, 13]. The triode extraction system is chosen as usually due to its advantages in generation of low energy and high current ion beams [14]. The magnetic field of DUHOCAMIS is a high bottle-shaped field with a remote magnetic region in the beam direction. It will seriously result in trajectory deviations of the extracted ions. We thus add a magnetic shield to reduce the edge-field and an electric deflector to minimize the beam deviations [12]. Fig. 3 shows this magnetic shield can effectively reduce the edge-field and simultaneously keep a high value in the discharge chamber. The analyzing magnet (9) with a bending radius of 0.35 m, is used to sort the ions having the determined charge-to-mass ratio to get into the target chamber; but only a few of ion spectra were measured for the source operated at lower magnetic field in the last experiments, because a focusing lens between the source and the analyzing magnet is being underway.

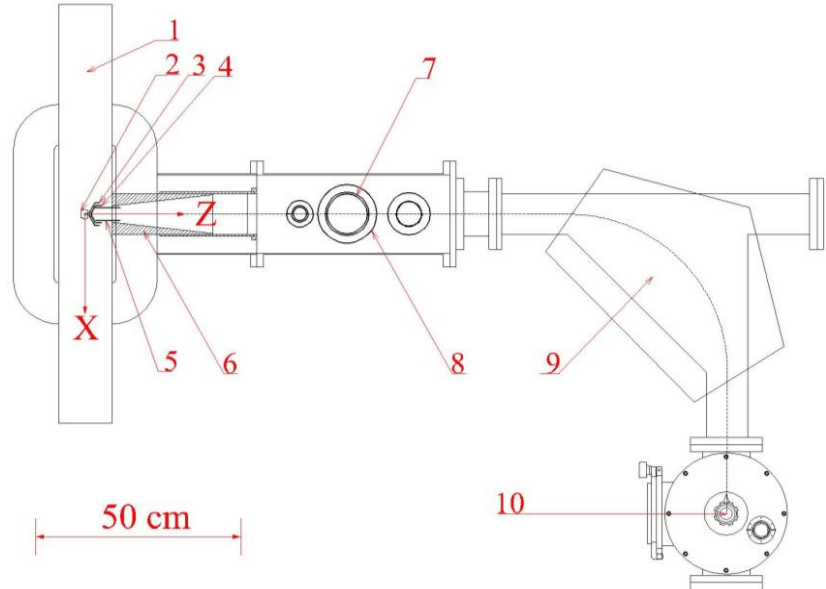

Fig. 2. Schematic diagram of the test bench for DUHOCAMIS: (1) Ion source magnet, (2-4) triode extraction system, (5) electrostatic deflector, (6) magnetic shield, (7) Faraday cup, (8) vacuum port, (9) analyzing magnet and (10) target chamber.

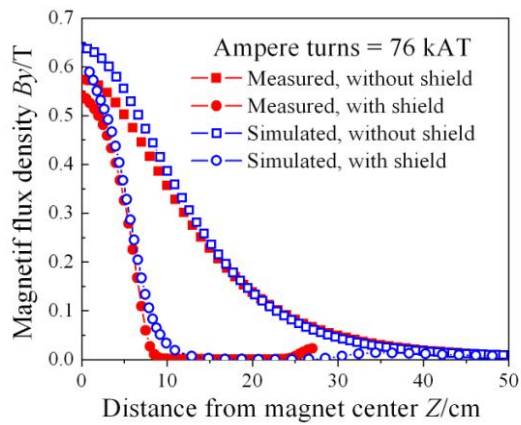

Fig. 3. Distributions of the magnetic flux density By along the beam line.

## 3. Experimental results and discussions*
### 3.1 Effect of cathode heating power on arc characteristics

With the conditions of a magnetic flux density of 0.38 T, a constant Argon flow of 1.4 cc/min, and a pulsed arc power supply with a pulse width of 1 ms & a frequency of 10 Hz, the discharge curves (V-I) of the source, i.e., the voltage-current characteristics were measured. Two typical curves corresponding to two cathode heating powers of 0.8 kW and 0.9 kW, respectively, are shown in Fig.4. The results indicate that different heating powers create different discharge characteristics: the higher power getting to the higher arc current with the lower curve slope. It means that the cathode heating power is sensitive to discharge, and the higher heating power is required to get higher arc current; in the mean time, the heating power should be stabilized enough. This experiment demonstrates that the slope of the V-I curves and the discharge current can be adjusted smoothly and stabilized simply by the indirectly heated cathode; it must be better to get multiply charged ions because of the accurate discharge parameters to be guaranteed probably [7].

* In all of the following experiments, the hollow sputter cathode was made of copper (Cu).

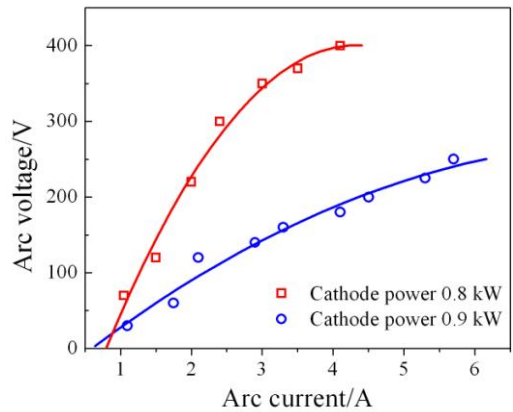

Fig.4. Effects of cathode heating power on discharge.

**3.2 Effect of magnetic flux density on arc characteristics**

Under the conditions: an applied cathode heating power of about 0.8 kW, an Argon flow of 1.4 cc/min, and an arc power supply with a pulse width of 1 ms & a frequency of 10 Hz, a series of discharge curves (V-I curves) were measured in a wide range of magnetic flux density from 0.13 T to 0.52 T. They are shown in Fig. 5.

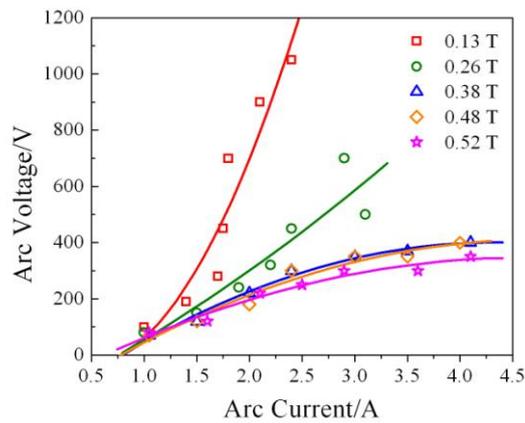

Fig. 5. Arc characteristics with different magnetic flux densities.

The results indicate that all the discharge currents monotonously increase with the increased arc voltages under a determined magnetic flux density; also the discharge currents increase with the increased magnetic fields for a determined arc voltage. It is interesting in slope of V-I curves, because the slope, i.e. dV/dI or dI/dV, shows the specific resistance or the specific conductivity in the plasma, respectively. From Fig. 5 we can see that the slope of the V-I curves can be varied in a wide range by changing the magnetic flux density; and a low slope is required for the high arc current. In fact, these V-I curves in Fig. 5 could be divided into three kinds of curves or three regions of magnetic flux density: resistance-stabilized, low-arc-current and low-arc-voltage regions; i.e., 1) the slope of V-I curves almost is a constant. The V-I curve is as a straight line around some magnetic field $B_s$ of about 0.26 T; 2) The slope is increased with the increased arc current in the left side of the $B_s$ field; 3) In the right side of the $B_s$ field, the slope is decreased with the increased arc current, and then gradually to zero. In addition, from Fig. 5 we can see that high arc currents probably could be obtained in the third region field only, for example, a field higher than 0.38 T, which is most interesting region to be investigated. It demonstrates that there exists a much stronger and sensitive magnetic effect to the DUHOCAMIS. Thus, we can say that the magnetic field is so important for the DUHOCAMIS, in which is not only the field geometry, but also the flux density.

**3.3 Ion spectra**

With an extraction voltage of 20 kV and a constant Argon flow of 0.7 cc/min, we measured some ion spectra of the DUHOCAMIS in the target chamber by the analyzing magnet. Fig. 6 presents the Ar-Cu ion spectra of the source operating in two different magnetic flux densities for discharge. It shows that the beam current ratios of $Cu^+/Ar^+$ and $Cu^{2+}/Ar^{2+}$ increase from 0.3 to 0.7 and 0.1 to 0.8, respectively, when the magnetic flux density of the source increases from 0.13 T to 0.26 T, also the $Cu^{2+}$ content became 2.2 times higher. The primary results confirm that the production of metallic ions would be much increased than gas ions with the increased magnetic flux density of the source.

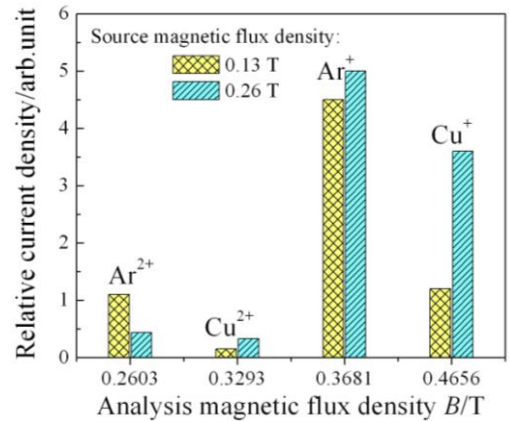

Fig. 6. Ar-Cu ion spectra.

**3.4 Comparison between DHCD and PIG discharge mode**

The DUHOCAMIS can change to PIG discharge mode from DHCD mode easily by the switch K shown in Fig. 1. So, with the same geometry and comparable discharge conditions, the following discharge experiments for PIG discharge mode have been done similarly for DHCD mode shown in Fig. 5. The results are shown in Fig. 7. It shows these V-I curves similar as in Fig. 5; but there is no "low-arc-voltage curves" appeared in Fig. 7, though they were measured in the same range of magnetic flux density from 0.13 T to 0.52 T. On the contrary, all kinds of curves in Fig. 7 could be found in DHCD mode. Meanwhile we can see that the maximal current obtained in Fig. 7 is much lower than that in Fig. 5. It means that the magnetic effects in DHCD mode are much stronger and more sensitive than that in PIG discharge mode.

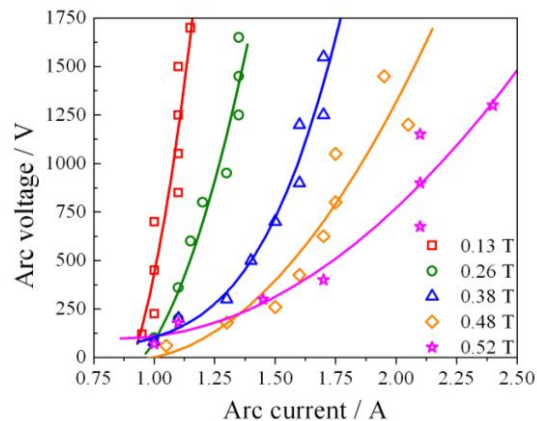

Fig. 7. Discharge curves for PIG mode.

To understand the differences between DHCD and PIG discharge mode of the source, we try to compare their discharge curves in more detail. Let's give a definition of "Factor of DHCD" ( $F_D$ in simplified form), that is $F_D == I_{DHCD} / I_{PIG}$, in which $I_{DHCD}$ and $I_{PIG}$ is the arc current in DHCD and PIG discharge mode, respectively, and they were taken at a discharge condition for the same arc voltage with the same magnetic flux density. The dependence of $F_D$ on the "arc voltage" and the magnetic field shows in Fig. 8, where the slope's trends of curves are very similar as in Fig. 5. It means that there also are three kinds of slope's trend: low-arc-voltage, resistance-stabilized, and low-arc-current curves. In Fig. 8 most of $F_D$ factor are higher than 2 except to a field 0.13 T. It demonstrates that the higher field gets the higher $F_D$, i.e. the magnetic effects in DHCD mode is much stronger than that in PIG discharge mode. Specially, when the field By was ≥ 0.38T in DHCD mode, the discharge worked in the low-arc-voltage region with a high current, i.e., the discharge was much stronger than that in PIG discharge mode at a field even more. The reason may be that there was a highly magnetized hollow cathode sputtering metal plasma in DHCD mode, but it was impossible to occur in PIG discharge mode. It might be true; there is a strong interaction between the sputtering metal plasma and the high magnetic bottle-shaped field. Thus, by comparison of discharge curves between DHCD and PIG discharge mode, it's clearer that the unique properties of the DUHOCAMIS would prominently demonstrate in operating at a high magnetic bottle-shaped field to deliver high metal ion beams.

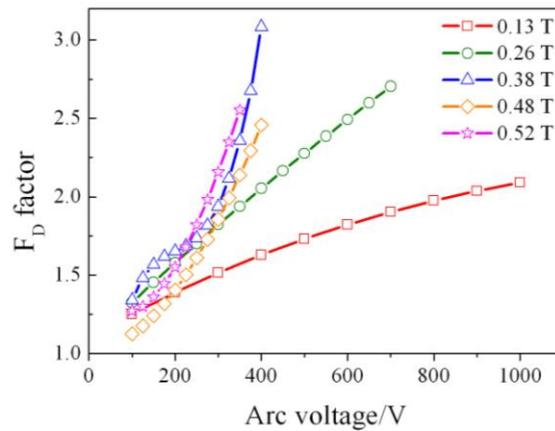

Fig.8. Dependences of $F_D$ factor on arc voltage and magnetic field.

## 4. Conclusions

A new test bench for the DUHOCAMIS with a high magnetic bottle-shaped field up to 0.6 T has been built, on which have been made a series of experiments in connection with discharge characteristics in a range of magnetic fields from 0.13 T to 0.52 T.

The experiments with discharge curves have demonstrated that the magnetic flux densities are very sensitive to the discharge behavior: discharge curves and ion spectra; the slope of discharge curves in a very wide range can be controlled by changing the magnetic field, as well as regulated by adjusting cathode heating power, independently. In comparison of discharge characteristics between DHCD and PIG discharge mode, it was found a much stronger magnetic effect occurred on DHCD mode, especially in a high magnetic field of about up to 0.4 T. It might be true; there is a strong interaction between the sputtering metal plasma and the high magnetic bottle-shaped field.

Thus, the importance of magnetic field for the DUHOCAMIS may be established not only by its special bottle-shaped structure, but also by its enough high flux density. The unique property for the DUHOCAMIS would be prominently demonstrated in operating at a high magnetic bottle-shaped field, from that it would be expected to get ion beams with high current, high charge state and high metal ion content.

*Dr. Michael Müller was once one cooperator for this project. He came to PKU three times, contributed his full experience of Dual HCD Ion Source and provided the ion source geometry. The authors wish to express the depth of their gratitude to Dr. Michael Müller.*

*The authors also would like to thank Drs. Peter Spädtke and Klaus Tinschert for their kind constant discussions from GSI-Darmstadt, Germany.*